# Broadband inverted T-shaped seismic metamaterial

Yi Zeng[1, 2], Shu-Yan Zhang[3], Hong-Tao Zhou[1], Yan-Feng Wang[1], Liyun Cao[2], Yifan Zhu[2], Qiu-Jiao Du[4], Badreddine Assouar[2, *] and Yue-Sheng Wang[1, *]

[1]*Department of Mechanics, School of Mechanical Engineering, Tianjin University, Tianjin 300350, China*

[2] *Institut Jean Lamour, CNRS, University de Lorraine, Nancy 54506, France*

[3] *Institute of Engineering Mechanics, Beijing Jiaotong University, Beijing 100044, China*

[4] *School of Mathematics and Physics, China University of Geosciences, Wuhan 430074, China*

\* Electronic mail: badreddine.assouar@univ-lorraine.fr (B. Assouar), yswang@tju.edu.cn (Y.-S. Wang).

**Abstract**

Seismic metamaterials (SMs) are expected to assist or replace traditional isolation systems owing to their strong attenuation of seismic waves. In this paper, a one-dimensional inverted T-shaped SM (1D ITSM) with an ultra-wide first bandgap (FBG) is proposed. The complex band structures are calculated to analyze the wave characteristics of the surface waves in the SMs. We find that the FBG of the 1D ITSM is composed of two parts; part 1 with surface evanescent waves and part 2 with no surface modes. Similar results are found in the complex band structure of the FBG of the SM consisting of periodically arranged pillars and substrate. The propagation of seismic surface waves in the 1D ITSM is different in these two frequency ranges of the FBG. In part 1, the seismic surface waves are significantly attenuated in the 1D ITSM, while in part 2, the surface waves are converted into bulk waves. Finally, the ultra-wide FBG is verified by using a kind of the two-dimensional

ITSM in large-scale field experiments.

**Kay words:** seismic metamaterials, surface waves, inverted T-shaped, first bandgap, relative bandwidth, large-scale field experiments

# 1. Introduction

Metamaterials[1-6] are a kind of composite material composed of periodically arranged artificial unit cells. One of the most significant characteristics of metamaterials is the bandgap (BG).[7-11] In the frequency range of the BG, waves cannot propagate in the metamaterials. This characteristic is used by physicists to control the propagation of seismic waves, thereby preventing casualties and property damage caused by the destruction of buildings. This kind of metamaterial used to control seismic waves is called seismic metamaterial (SM).[12-16] The SM can be used to protect critical infrastructures and most buildings located in the mid- and far-fields of the earthquake source. It is worth noting that the natural frequency of these buildings is below 20Hz.[12,17] Therefore, the realization of the SM with ultra-low BGs is an important step to push the SM to practical application. The conjecture of the SM was proposed after the experiment of the phononic crystal in marble in 1999.[7] However, a large amount of research on the SMs began in 2014. In this year, cylindrical holes arranged periodically in soil were experimentally proved to attenuate seismic waves at 50 Hz.[14] These large-scale experiments on the SM have shown that it has unlimited potential in controlling seismic waves. Later, researchers changed the shape of the holes, and filled different materials into the holes to obtain the SMs with a wide BG.[18-20] However, the structure of this kind of SM based on the Bragg scattering mechanism is very huge to control ultra-low-frequencies seismic waves, which makes it difficult for practical application.

Fortunately, the SMs based on local resonance can solve this problem, which can use small structures to control long waves.[21-24] Therefore, most of these SMs are also called subwavelength SMs.[25-27] The subwavelength SMs can be roughly divided into two categories. The first kind of subwavelength SMs can also be called the underground barrier of locally resonant metamaterial.[15,28-31] This kind of underground SMs with small scale is based on resonance characteristics to attenuate long-wavelength seismic waves.[21] However, their BGs are generally narrow. And the depth of the SMs burial into the ground should be roughly equal to twice the wavelength of the seismic surface wave to better attenuate the seismic surface wave. Because the energy of surface waves is mainly concentrated in this depth range.[32] The second kind of subwavelength SMs are usually composed of periodic pillar-like structure and substrate, which can be called the pillar-like SM (PSM). A major characteristic of the PSM is that it mainly attenuates the propagation of seismic surface waves,[33,34] especially in the low-frequency range of the first BG (FBG), and even if the PSM only has three rows.[34,35] The PSM is found because seismic surface waves in a part of the frequency range cannot pass through forest.[34] I-shaped,[35] fractal structure [36] and Matrayoshka-like [37] PSMs are proposed to obtain a wider complete BG below the sound cone. However, the relative bandwidth of the FBG is currently too small to meet the needs of practical applications.

In this paper, a one-dimensional inverted T-shaped SM (1D ITSM) composed of pillars and plates with ultra-wide FBG is proposed. The ITSM is just a simple modification on the basis of the PSM: a plate is added at the bottom of each pillar. The FBGs of the two different SMs are compared. The effects of the ITSM's geometric and material parameters on the FBG are discussed. The complex band structure of the ITSM for surface waves is calculated and compared with that of the PSM. The propagation properties of seismic surface waves in the ITSM are also given. Finally, we propose a kind of the two-dimensional (2D) ITSM with the ultra-wide FBG.

## 2. Model of the 1D ITSM

In this section, the classical PSM and the proposed 1D ITSM are compared. As shown in Fig. 1, the PSM and the ITSM have exactly the same steel pillar, i.e., $l_2 = l$, $w_2 = w$. But the bottom of every pillar in the ITSM is clamped by a steel plate marked as gray part in Fig. 1(b). The geometric and material parameters of the unit cells are given in Tables 1 and 2, respectively. In numerical simulation, the left and right boundaries of the substrate of the unit cell are set as Bloch periodic boundaries, and the bottom of the substrate is set as a fixed boundary condition to avoid surface modes on the bottom.[19,20,37] It is worth noting that the calculation results will converge well at ultra-low frequency when the geometric parameter $H$ is enough large. In this paper, the COMSOL Multiphysics is used to calculate the band structures and the propagation of Rayleigh waves in the SMs.

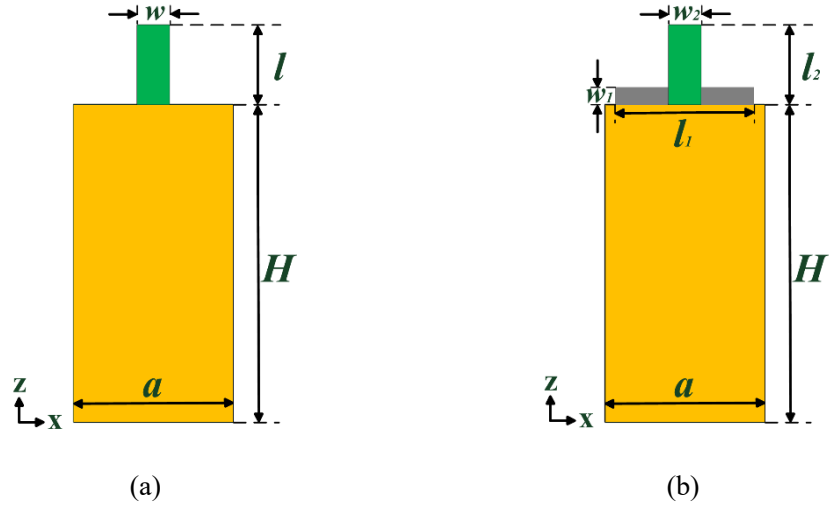

(a)        (b)

Figure 1: The unit cells of (a) the PSM and (b) the 1D ITSM.

Table 1: The geometric parameter of the PSM and the 1D ITSM

| $a$ | $H$ | $l_1$ | $w_1$ | $l_2(l)$ | $w_2(w)$ |
|---|---|---|---|---|---|
| 1.5 m | 300 $a$ | 0.95 $a$ | 0.05 $a$ | 1.4285 $a$ | 0.07 $a$ |

Table 2: The material parameters used in this paper[18,28,30]

| Material | Density (kg/m$^3$) | Young's modulus (Pa) | Poisson's ratio |
|---|---|---|---|
| Steel | 7784 | $2.07 \times 10^{11}$ | 0.3 |
| Soil | 1800 | $2 \times 10^7$ | 0.3 |

## 3. Results and Discussion of the 1D ITSM

### 3.1 Classical band structure

Figures 2 (a - b) show the classical band structures of the PSM and the ITSM, respectively. The classical band structures are calculated by using the Solid Mechanics module of COMSOL. The frequency range of the FBG of the PSM is from 0.72 Hz to 0.80 Hz. For the ITSM, it is from 6.7 Hz to 17.2 Hz. Although the FBG of the PSM is very low, it is extremely narrow (relative bandwidth is about 0.1) and thus unsuitable for seismic shielding. In contrast, although the minimum value of the FBG of the ITSM is 6.7 Hz, its relative bandwidth can reach an astonishing value as 0.88. We believe that this way of abandoning the BG with ultra-low frequency in exchange for the ultra-wide BG is desirable in the practical application of SMs.

The vibration modes of the PSM and the ITSM at marked points are illustrated in Figs. 2 (c - d). The direction of the arrows represents the direction of movement of the particles, and the color indicates the normalized total displacement. It is easy to find that the vibration modes of the lower, upper boundary of the two FBGs (points $A_1$ and $A_2$, points $B_1$ and $B_2$) are similar, respectively. This indicates that the generation mechanism of these two FBGs is also similar. On the first bands of two SMs, the maximum displacement appears on the top of the steel structure. On the second bands of two SMs, the displacement is on the entire steel structure and on the surface of substrate. The energy is evenly distributed on the entire structure. The pillar or the inverted T-shaped structure (ITS) can be regarded as a whole at this time. This shows that the structure of the ITS leads to a qualitative leap of the FBG.

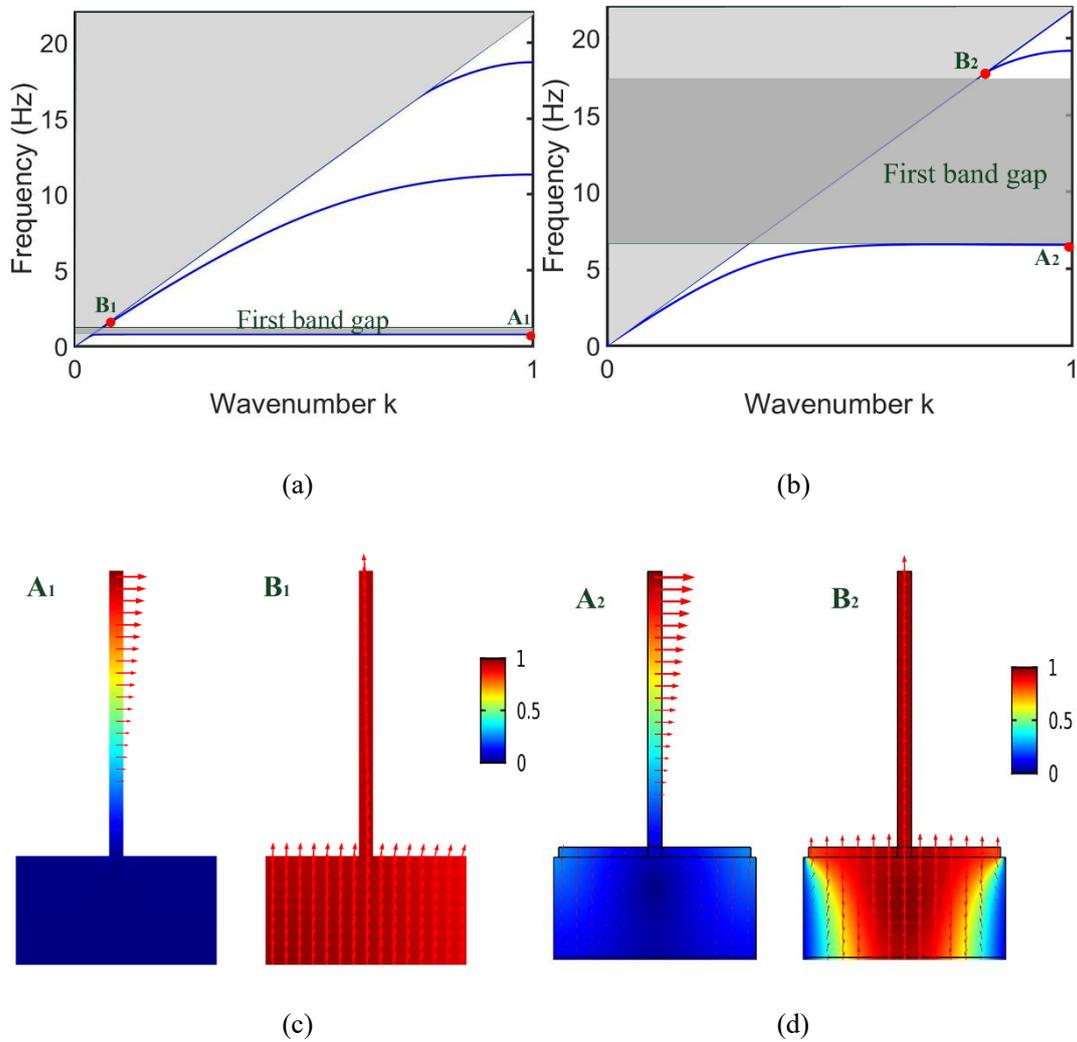

Figure 2: The classical band structure of (a) the PSM and (b) the 1D ITSM, the vibration modes of (c) the PSM and (d) the 1D ITSM at the points marked the classical band structures.

### 3.2 Effects of geometric parameters

After comparing the band structures of the PSM and the ITSM, it is obvious that the appearance of the plate at the bottom of the pillar makes a huge change in the frequency range and relative bandwidth of the FBG. In this section, the effects of geometric parameters of the plate on the FBG are calculated, when the pillar is unchanged. In addition, the effect of the height of the pillar on the FBG is calculated to obtain the optimal geometric parameters of the ITSM, when the total volume of the pillar (i.e., $l_2 \times w_2$) is constant.

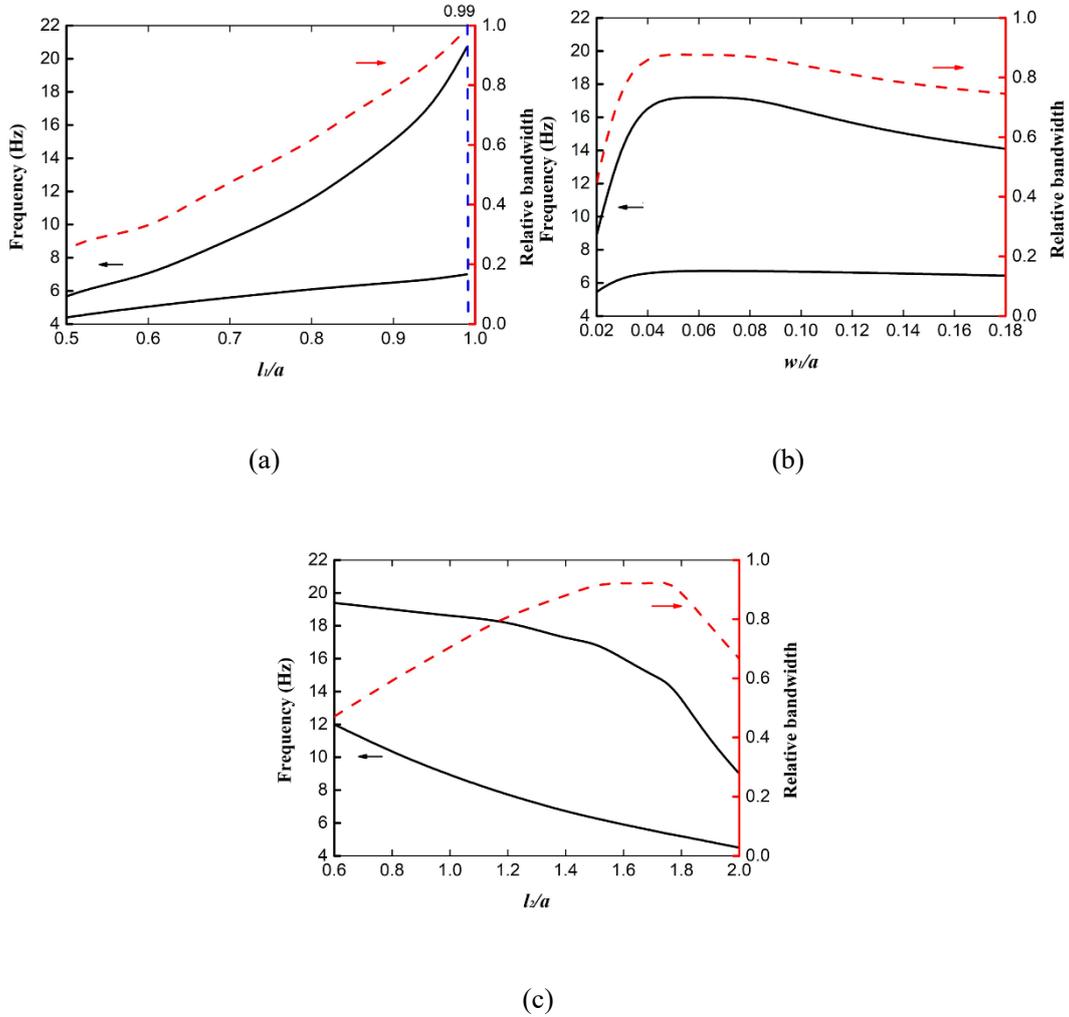

(a)   (b)

(c)

Figure 3: Variation of the upper and lower frequency boundaries, and the relative bandwidth of the FBG with the ratio of the geometric parameters (a) $l_1$, (b) $w_1$ and (c) $l_2$ to lattice constant $a$. The black solid line is the upper and lower boundaries of the FBG, and the red dashed line is the relative bandwidth.

Figures 3(a - b) show that the variation of geometric parameters of the plate with the position and relative bandwidth of the FBG. As shown in Fig. 3(a), when the length $l_1$ of the plate increases from 0.5$a$ to 0.99$a$, the lower boundary of the FBG slowly increases from 4.4 Hz to 7.0 Hz, and the upper boundary increases rapidly from 5.7 Hz to 20.7 Hz. Therefore, the relative bandwidth of the FBG increases from 0.27 to 0.99. As shown in Fig. 3(b), when the width of the plate increases from 0.02$a$ to 0.18$a$, the lower boundary of the FBG first grows slowly from 5.5 Hz to 6.7 Hz, and then slowly drops to 6.4 Hz. However, the upper boundary of the FBG increases rapidly from 9.0 Hz to 17.2 Hz, and then slowly decreases to 13.7 Hz. Therefore, the relative bandwidth of the FBG increases rapidly from 0.47 to 0.88, and then slowly drops to 0.73.

Figure 3(c) shows the effect of the pillar height on the FBG. When the pillar height continues to increase from 0.6$a$ to 1.7$a$, the upper and lower boundaries of the FBG continue to decline, and its relative bandwidth continues to increase. But when the pillar height is greater than 1.8$a$, the upper boundary of the FBG drops rapidly, leading to the decrease of the relative bandwidth. Therefore, when the pillar height is about 1.7$a$, the relative bandwidth of the FBG of the ITSM is the largest, about 0.93.

The results show that the geometric parameters of the plate and pillar in the ITS have a very obvious effects on the relative bandwidth of the FBG. With the appearance of the plate and the increase in its length, although the center frequency of the FBG of the ITSM continues to increase, its relative bandwidth increases linearly. However, there is an optimal value for the thickness of the steel plate, about 0.05$a$. When the height of the pillar continues to increase within a certain range, the center frequency of the FBG continues to decrease, but the relative bandwidth continues to increase. However, when the pillar is too high and its width is too small ($l_2 \times w_2$ is constant), the decrease in the flexural resonance frequency of the pillar causes the relative bandwidth to decrease continuously.

## 3.3 Effects of material parameters

In this section, the influence of material parameters of the ITS on the FBG is analyzed. Material parameters of the ITS are swept only for qualitative comparison.[38] Only one parameter is changed at a time, and other parameters are consistent with those in Fig. 1(b).

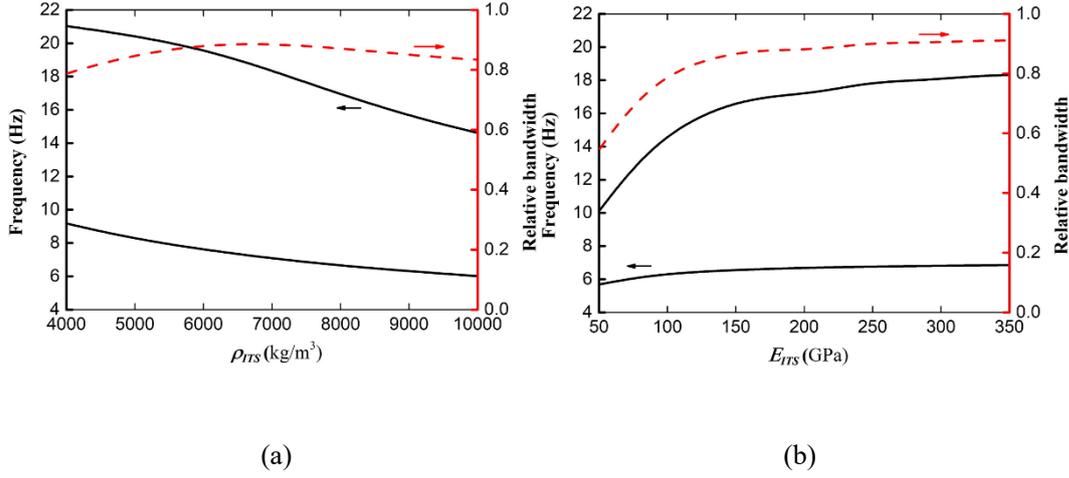

(a)                                                  (b)

Figure 4: Variation of the upper and lower frequency boundaries, and the relative bandwidth of the FBG (a) the mass density $\rho_{ITS}$ and (b) the Young's modulus $E_{ITS}$. The black solid line is the upper and lower boundaries of the FBG, and the red dashed line is the relative bandwidth.

The effect of mass density of the ITS ($\rho_{ITS}$) on the FBG is shown in Fig. 4(a). When $\rho_{ITS}$ increases from 4000 kg/m³ to 10000 kg/m³, the lower boundary of the FBG decreases from 9.1 Hz to 6.0 Hz; and the upper boundary reduces from 21.0 Hz to 14.5 Hz. The center frequency of the FBG keeps dropping. However, the relative bandwidth first increases and then decreases, and its maximum value is about 0.9 when $\rho_{ITS}$ is about 7000 kg/m³. The effect of Young's modulus of the ITS ($E_{ITS}$) on the FBG is shown in Fig. 4(b). When $E_{ITS}$ increases from 50 GPa to 350 GPa, the lower boundary increases slightly. When $E_{ITS}$ increases from 50 GPa to 180 GPa, the upper boundary rapidly increases from 10.0 Hz to 17.0 Hz. Finally, the frequency slowly increases to 18.2 Hz, when $E_{ITS}$ increases from 180 GPa to 350 GPa. Therefore, the relative bandwidth rapidly increases from 0.54 to 0.88, and it has just little change when $E_{ITS}$ is larger than 180 GPa. These results show the ITSM has the FBG with largest relative bandwidth when $\rho_{ITS}$ is about 7000 kg/m³ and $E_{ITS}$ is larger than 180 GPa. In reality, steel is the best choice for the ITS.

**3.4 Complex band structure**

In order to deeply explore the wide FBG of the ITSM, the complex band structures shown in Fig. 5(a), are calculated by using the PDE module of the COMSOL. In order to draw only the surface modes in the complex band structure, the parameter ξ is introduced, i.e.

$$\xi = \int_{S(2\lambda)} |\boldsymbol{u}|ds \,/\, \int_{S(H)} |\boldsymbol{u}|ds \tag{1}$$

that is the ratio of the integral of the displacement ($\boldsymbol{u}$) in the range of the Rayleigh wave wavelength with a depth of twice ($2\lambda$) to the integral of the displacement over the entire depth ($H$) of the substrate for the ITSM. In this paper, surface mode is defined as ξ > 0.9. In Fig. 5(a), the color bar on the far right represents the value of ξ. The black dotted line is the boundary of the sound cone.

In the complex band structure of the ITSM, the FBG can be divided into two completely different parts: part 1 with surface evanescent waves (light gray); part 2 with no surface mode (heavy gray). From the real part of the complex band structure of the ITSM, there is a surface band in the FBG in part 1, with a frequency range from 6.7 Hz to 11.0 Hz. This band partially overlaps with the first band at about 6.7 Hz. In the imaginary part of the surface band in part 1 are all greater than zero. They are surface evanescent waves in this frequency range and cannot be displayed in the classical band structure. The vibration mode at the end of the band is plotted. It can be found that most of the vibration exists in the substrate and decreases with increasing depth. We speculate that as the frequency increases, most of the vibration is transferred to the substrate thus the surface mode disappears. In part 2 with the frequency range of 11.0 Hz to 17.2 Hz, there is no surface band in the complex band structure of ITSM. Similar phenomenon is currently only found in the inverse dispersion SMs (Appendixes A and B). We speculate that in this frequency range, the steel structure can be equivalent to a layer of uniform medium on the substrate due to its longitudinal resonance (vibration perpendicular to the surface of the substrate). The layered media structure composed of this medium and the substrate has an inverse dispersion BG. In this frequency range of the BG, the surface waves will be converted into bulk waves (Appendixes B).

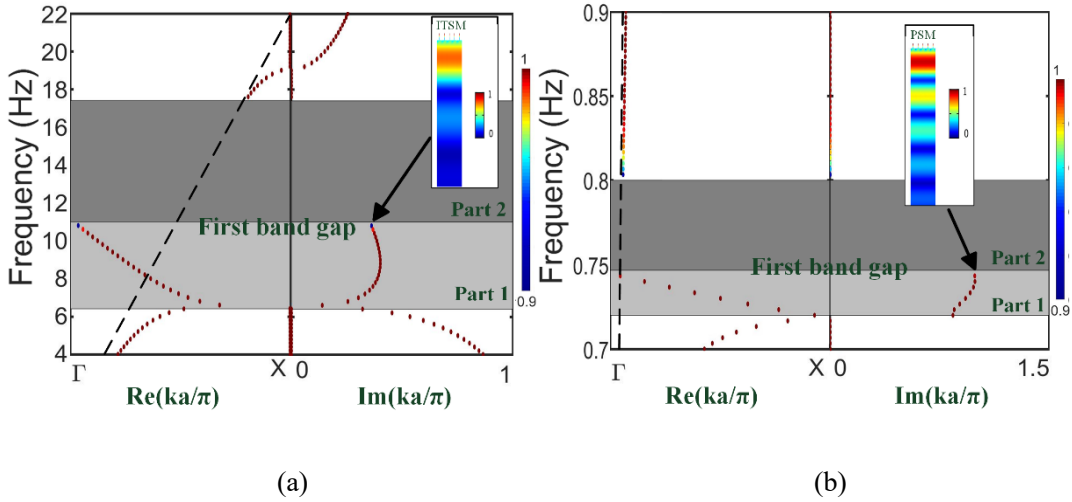

(a)            (b)

Figure 5: The complex band structures of (a) the 1D ITSM and (b) the PSM. The color bars from 0.9 to 1 on the far right represent the value of $\xi$. The color bars from 0 to 1 embed in pictures represent the displacement in the vibrating modes.

Similarly, as shown in Fig. 5(b), as a comparison, the complex band structure of the FBG of the PSM shown in Fig. 1(a) is calculated. It is easy to find that the situation in these two FBGs is almost identical. So, we believe that the formation mechanism of these two FBGs is almost the same.

### 3.5 Frequency domain analysis

The transmission spectrum is calculated in this section. The finite model used for calculation is shown in Fig. 6(a), and consisting of two parts: the substrate represented by the yellow part, and the periodic ITSs represented by the gray part and the green part. The material and geometric parameters of the ITS are consistent with those in Fig. 1(b). The boundaries of the left, right, and bottom of the substrate are set as low reflection boundary condition to eliminate the reflection of elastic waves.[19,20] The number of rows of the ITSM is 10. Point A is set as the source to generate Rayleigh waves. The polarization direction of the source is the *xz* direction.[26,39,40] Accordingly, the data is collected at point B from the right side of the ITSM. Similarly, in order to obtain an accurate transmission spectrum at a sufficiently low-frequency range, the height of the substrate is set as $H = 300a$. The acceleration at point B is collected with and without the ITSM. When there is the ITSM, the acceleration at point B is $A_1$. When there is no ITSM, the acceleration at point B is $A_0$. The transmission is defined as $T = 20 \times \log_{10}(A_1/A_0)$.

As shown in Fig. 6(b), the ITSM has a very significant attenuation effect on the Rayleigh waves in the frequency range of the FBG. There is a huge attenuation zone from 6.7 Hz to 17.2 Hz, consistent with the range of the FBG in Fig. 2(b). This result shows that the ITSM can drastically attenuate the Rayleigh waves at a sufficiently low frequency and in a wide range. In addition, it is easy to find that in the FBG's two different parts identified in complex band structure (Fig. 5(a)), the attenuation effects of the ITSM on Rayleigh waves are completely different. In part 1, the attenuation effect is obviously better than that in part 2, because there are evanescent surface waves. In part 1, the larger the value of the imaginary part, the stronger the attenuation effect of the ITSM on Rayleigh waves.

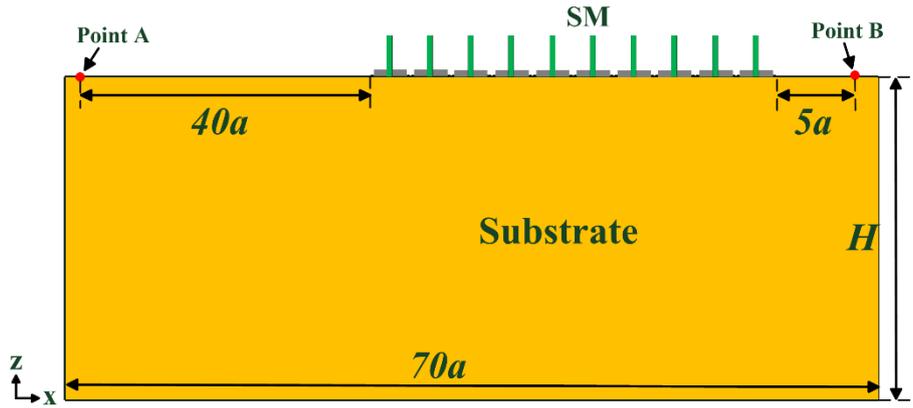

(a)

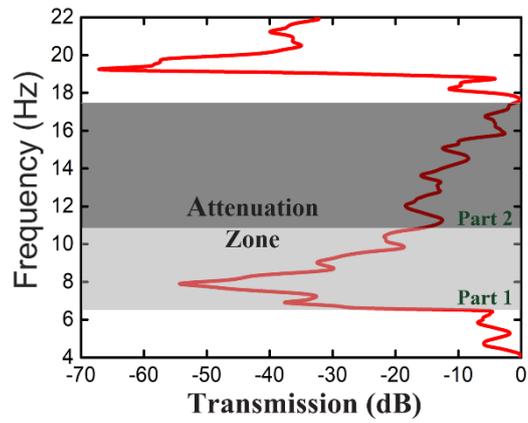

(b)

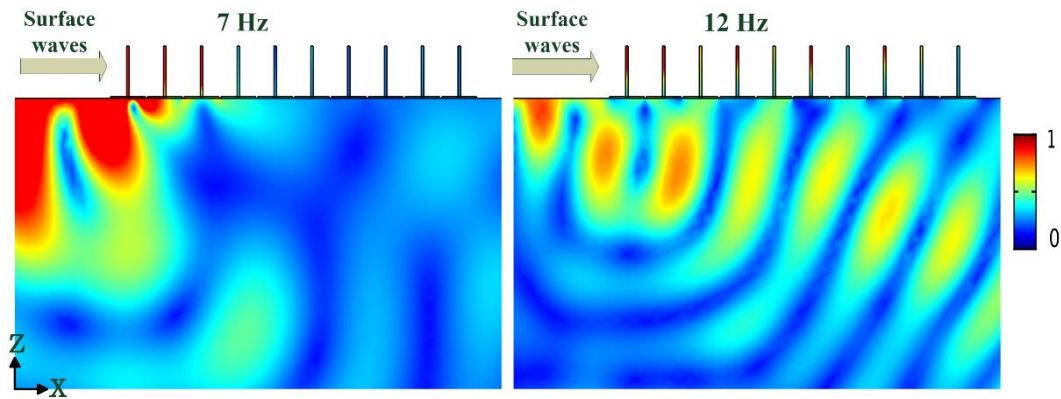

(c)

Figure 6: (a) Schematic model of the 1D ITSM with 10 unit cells for transmission calculations. (b) Transmission spectrum of the ITSM at the frequency range from 4 Hz to 22 Hz. (c) Distribution of displacement amplitude in the ITSM at 7 Hz and 12 Hz.

As shown in Fig. 6(c), the distribution of displacement amplitude in the ITSM is given at two different but representative frequencies in the FBG: 7.0 Hz and 12.0 Hz. In the frequency range of part 1, for example, at 7.0 Hz, the violent resonance of the ITSM causes the surface waves to be significantly attenuated and almost unable to propagate to the third unit cell. In the frequency range of part 2, for example, at 12.0 Hz, after the surface waves enter the ITSM, we can observe that the surface waves are converted into bulk waves. These two results are easily reminiscent of the literature, [33] where surface waves are incident from two ends of the metawedge. When the surface wave is incident from the short edge, the surface waves first encounter part 1 of the FBG of the shorter PSM. So, the surface wave is quickly attenuated because of the evanescent waves. When the surface wave is incident from the taller end, the surface waves first encounter part 2 of the FBG of the taller PSM. So, the surface waves are converted into bulk waves because of the inverse-dispersion effect (Appendixes A). This also indirectly proves that the formation mechanism of the FBG of the PSM and the ITSM is almost the same.

**4. The 2D ITSM and large-scale field experiments**

For the SM, a 1D periodic model is a general choice for exploring mechanism of the BGs.[20,26,32] However, in practical applications, the design of the 2D periodic SM is more important and necessary.[18,37,41] In addition, due to some uncontrollable factors in actual situations, such as the difficulty of achieving perfect continuity between different materials, large-scale field experiments are still needed to verify the SMs with ultra-wide and low-frequency bandgaps for surface waves. Therefore, based on the design concept and the mechanism of the above-mentioned 1D ITSM, a kind of 2D ITSM and a large-scale field experiments are designed for isolating Rayleigh waves in this section.

## 4.1 Designs of the 2D ITSM and experimental setup

Fig. 7(a) shows the unit cell of the 2D ITSM designed according to the structure of the 1D ITSM. In order to facilitate field experiments and attenuate seismic surface waves in the lower frequency range, the lattice constant $a_1$ of the unit cell is set as 0.25m. And the unit cell is composed of a thumbtack-shaped structure (purple part) and a soil substrate (gray part). Among them, the thumbtack-shaped structure (TS) is composed of a steel tube and a steel plate. The geometric parameters of the unit cell are shown in Table 3. As shown in Fig. 7(b), the central garden of the Department of Mechanics on the Beiyangyuan Campus of Tianjin University (China) was selected for a large-scale field experiments in August 2020. The geometric parameters of the TS used in large-scale filed experiments are consistent with those in Fig. 7(a). Among them, the steel tube and steel plate are rigidly connected by high-strength bolts. 25 identical TSs are customized and periodically arranged on the substrate.

Table 3: The geometric parameters of the unit cell in Fig. 7(a)

| $a_1$ | b | t | h | D | d |
|---|---|---|---|---|---|
| 0.25 m | 0.2375 m | 0.012 m | 0.75 m | 0.051 m | 0.041 m |

Taking into account the viscoelasticity and non-uniformity of the soil, we use the controlled variable method: using two sensors (uT, A21D100) with the same distance to the source to measure the acceleration of the two points (points A and B shown in Fig. 7(b)). The vibration source is generated by the rammer (Mingtu Mechanics, C90t) with an exciting force of 20kN. The data of two sensors is collected by the signal collector (uT, uT3604FS-ICP), and processed by the computer to obtain the response results at the two points. When the TSs are not placed in the field, the responses of the two points are almost the same when the rammer works. Therefore, this experiment can almost eliminate most uncontrollable factors, and only focus on the influence of the presence of artificial structures on the propagation of Rayleigh waves. The transmission of Rayleigh waves by the TSM is defined as $20*\log_{10}(A_A/A_B)$ [20], where $A_A$ and $A_B$ are the accelerations in the vertical direction of points A and B, respectively. The material parameters of the tests is referenced in these literatures [28,30], which is same as in Tables 2. It is worth noting that in the tests, the TS is simply placed on the substrate. But in the simulation mentioned above, the TSs and soil are perfectly continuous.

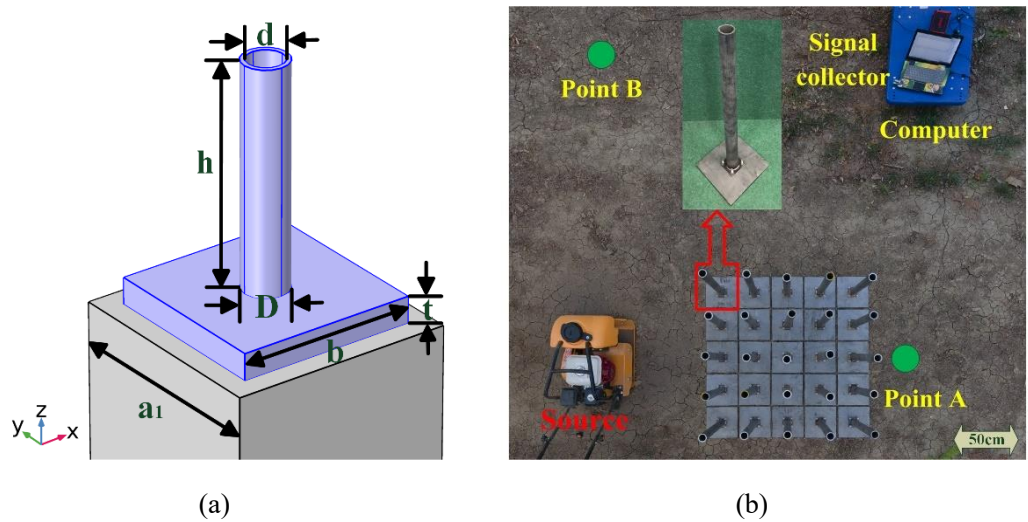

Figure 7: (a) Schematic of the unit cell and (b) top view of the experimental setup. Points A and B are signal acquisition points, and point B is the reference point.

## 4.2 Results of the 2D ITSM and field experiments

The band structure of the 2D ITSM in the ΓX direction is calculated, as shown in Fig. 8(a).[18,37,42] The dark gray part is the sound cone. All modes outside the sound cone are surface wave modes. The vibrating mode of the highest point of the first band is given in the insert of Fig. 8(a). It can be seen that the displacement is mainly concentrated on the top of the steel tube in $x$ direction. The vibrating mode of the second band exhibits also a displacement mainly concentrated on the top of the steel tube but in *y* direction. The surface waves mainly considered in this work are Rayleigh waves, whose displacement components on the surface only exist in the *x* and *z* directions [39]. Therefore, Rayleigh waves can propagate in the ITSM in the frequency range of the first band, but not the second band [36]. So, the bandgap of the ITSM for Rayleigh waves is in the range of 34-130 Hz, which is marked by the light gray part in Fig. 8(a).

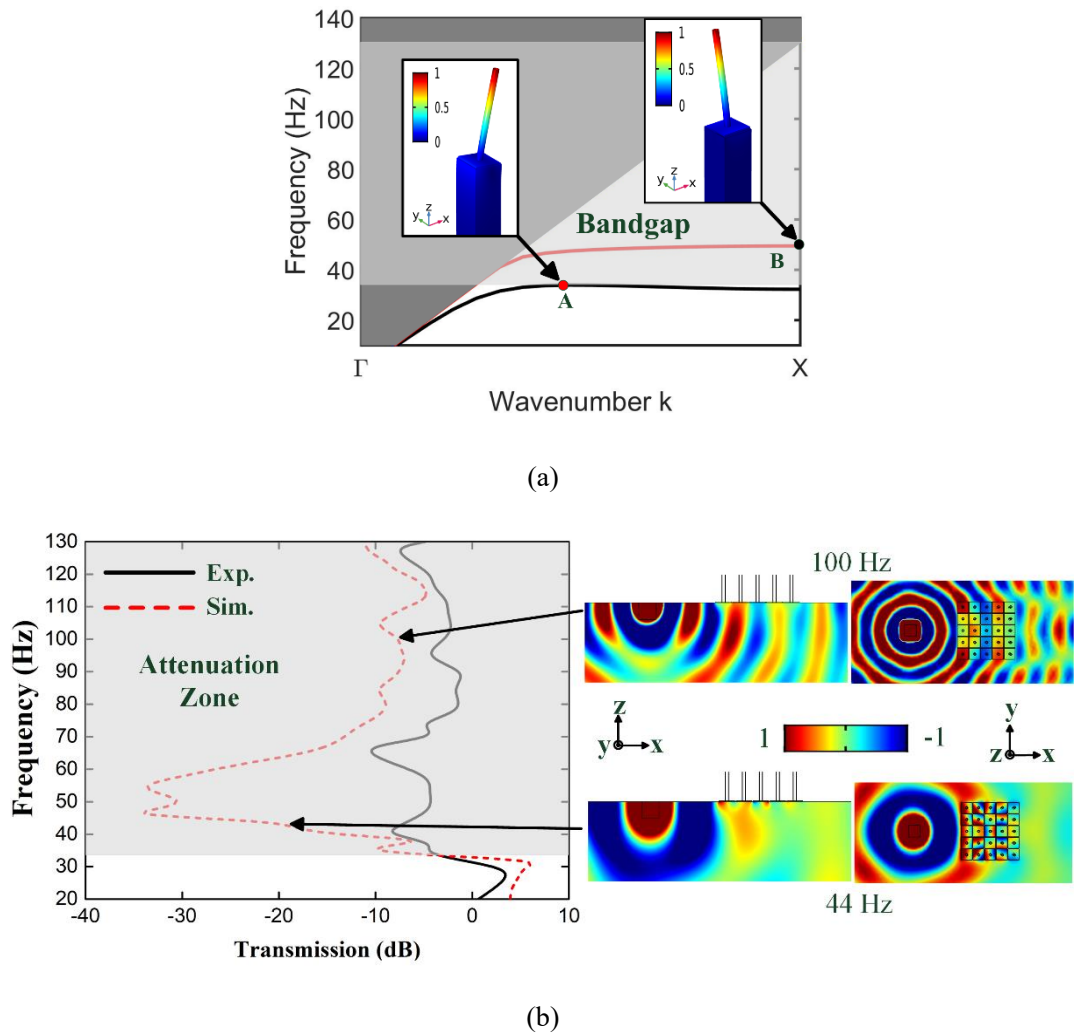

(a)

(b)

Figure 8: (a) the band structure in the ΓX direction of the first irreducible Brillouin zone of the 2D ITSM. (b) The transmission spectrum of the experimental and simulated results along ΓX direction through the 2D ITSM. The bandgap for Rayleigh waves and attenuation zone are all marked by the light gray part. And the cross-sectional and top views of simulated results of Rayleigh waves in the ITSM at 44 and 100 Hz are shown in the right side, respectively.

In Fig. 8(b), the black line and red dotted line are the experimental and simulation results, respectively. The model used in the simulation calculation is exactly the same as the experimental site. In the range of 34-130 Hz, Rayleigh waves are attenuated significantly. It can be found that the attenuation effect in the simulation is generally stronger than the result of the field test. This is due to the fact that the simple connection between the TSs and the soil in the field test is weaker than the perfect continuity in the simulation calculation. In addition, it is worth noting that in the bandgap, experimental and simulated results all show that the attenuation of Rayleigh waves is stronger at the band of 34-70 Hz than the one of 70-130 Hz. In the cross-sectional and top views of simulation results, it is easy to find that the propagation of Rayleigh waves in the ITSM is completely different at 44 and 100 Hz, respectively. (The propagation of Rayleigh waves without the ITSM is shown in Appendix C). This is consistent with the results in Fig. 6(c): the attenuation in the bandgap generated by the local resonance is stronger than that by inverse-dispersion characteristics when the rows of the unit cell are fewer (5 rows in this work).

## 5. Conclusion

In this research, an 1D ITSM with an ultra-wide FBG has been proposed by improving the pillar-like SM. With the total mass of the pillar unchanged, the relative bandwidth of the FBG of the ITSM is about 1 at maximum. This result is much better than that of the PSM. The effects of the geometrical and material parameters of the ITSM on the FBG have been also discussed. It is found that the appearance of the plate obviously increases the center frequency and relative bandwidth of the FBG of the ITSM. The complex band structure of the ITSM has been calculated to analyze the attenuation mechanism of surface waves. It can be found that the FBG of the ITSM is composed of two parts: part 1 with surface evanescent waves and part 2 with no surface mode. We also found that this result is consistent with the PSM. The propagation of seismic surface waves in the ITSM has been also investigated. It was found that the propagation modes of surface waves in the ITSM are completely different in different parts of the FBG. In the frequency range of part 1, the seismic surface waves are significantly attenuated and almost unable to propagate to the third unit cell in the ITSM. In the frequency range of part 2, the surface waves are converted into bulk waves. Finally, we have proposed a kind of the 2D ITSM with 25 unit cells in large-scale field experiments, which can attenuate Rayleigh waves in a ultra-wide frequency range. Moreover, the lattice constant in this work is much smaller than the wavelength of the surface waves in the ultra-low frequency range. A lower-frequency FBG can be obtained by appropriately increasing the lattice constant of the ITSM. This work not only provides new options for controlling seismic surface waves at ultra-low frequency, but also provides new design ideas for steering surface wave.

## Acknowledgments

This work has been supported by la Région Grand Est, the Institut CARNOT ICEEL, and the National Natural Science Foundation of China (NNSFC) under Grant Nos. 11702017, 11991031, 11991032, 12021002 and 41974059. The first author is grateful for the support of China Scholarship Council (CSC Grant No. 202006250084).

## Data availability

The data that support the findings of this study are available from the corresponding author upon reasonable request

**Appendix A: Inverse-dispersion effect**

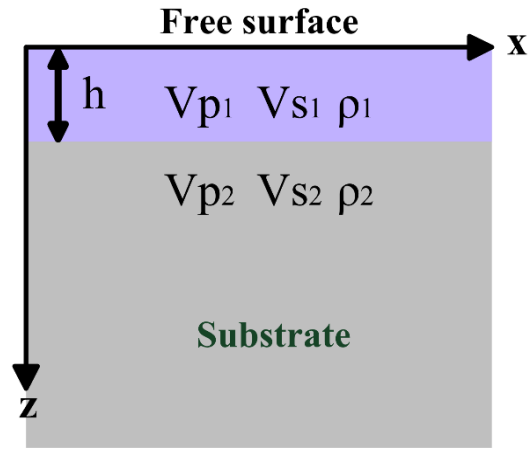

Figure 9: Layered media structure of a cover layer and a half-space substrate.

Figure 9 shows a layered model which consists of a homogeneous isotropic layer of thickness $h$ overlying a homogeneous half space. For the two-layered structure, the dispersion relation of the Rayleigh wave can be calculated using the transfer matrix method. [43,44] When the material parameters of the two-layered structure are identical, i.e., the homogeneous half-space structure, the velocity of the Rayleigh waves is constant. [27] When the material parameters of the two-layered structure are different, situations are complicated, and only two representative cases are discussed here: I ($V_{s2}>V_{r2}>V_{s1}>V_{r1}$) and II ($V_{s1}>V_{r1}>V_{s2}>V_{r2}$). The assumed parameters are shown in Table 4 and Table 5. Where $V_{p1}$, $V_{s1}$, $V_{r1}$, and $\rho_1$ ($V_{p2}$, $V_{s2}$, $V_{r2}$, and $\rho_2$) are the longitudinal wave velocity, shear wave velocity, Rayleigh wave velocity and density of the cover layer (substrate), respectively. The thickness of the cover layer is fixed at $h = 6$ m.

For case I, the dispersion curves of the Rayleigh waves of this two-layered layer structure are shown in Fig. 11(a). This is a very common dispersion curves in earth science. [45] There are multiple guided waves in the layered structure, and the maximum

Rayleigh wave velocity (phase velocity) is equal to the shear wave velocity of the substrate ($V_{s2}$), and the minimum Rayleigh wave velocity (phase velocity) is equal to the Rayleigh wave velocity of the cover layer ($Vr_1$). $V_{s2}$ and $Vr_1$ are underlined in Table 4. For any one of these guided waves, the velocity of the Rayleigh waves gradually decreases with increasing frequency and finally becomes flat.

Table 4: The assumed material parameters when $V_{s2}>Vr_2>V_{s1}>Vr_1$

|  | Vp | Vs | Vr | ρ |
|---|---|---|---|---|
| Cover layer | 1500 | 1000 | <u>893.1</u> | 1800 |
| Substrate | 2000 | <u>1400</u> | 1229.0 | 2300 |

Table 5: The assumed material parameters when $V_{s1}>Vr_1>V_{s2}>Vr_2$

|  | Vp | Vs | Vr | ρ |
|---|---|---|---|---|
| Cover layer | 2000 | 1400 | 1229.0 | 2300 |
| Substrate | 1000 | <u>800</u> | <u>636.7</u> | 1800 |

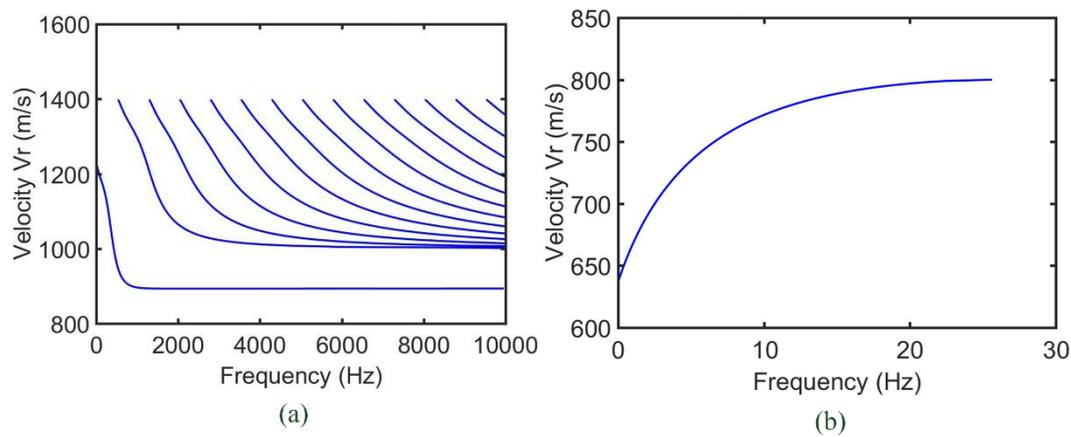

Figure 10: The dispersion curves of the Rayleigh waves of the two-layered layer structure when (a) $V_{s2}>Vr_2>V_{s1}>Vr_1$ and (b) $V_{s1}>Vr_1>V_{s2}>Vr_2$.

For case II, the dispersion curves of the Rayleigh waves of this two-layered layer structure are shown in Fig. 10(b). We can find the inverse-dispersion effect, [46] that is, the wave velocity (phase velocity) of the Rayleigh waves increases with increasing frequency. This layered structure has only a unique Rayleigh guided wave: the minimum Rayleigh wave velocity (phase velocity) is equal to the Rayleigh wave velocity of the substrate ($Vr_2$), and the maximum Rayleigh wave velocity (phase velocity) is equal to the shear wave velocity of the substrate ($Vs_2$). Also, $Vr_2$ and $Vs_2$ are underlined in Table 5. And the most interesting thing is that this Rayleigh guided wave has a cutoff frequency. When the frequency is higher than this cutoff frequency, about 25 Hz as shown in Fig. 10(b), there is no Rayleigh wave in such two-layered structure.

**Appendix B: A simple case for the BG based on inverse-dispersion effect**

Figure 11(a) shows the unit cell of the T-shaped SM (TSM) composed of T-shaped steel and substrate, and the substrate is the soil. The boundaries identified by the red lines in the Fig. 11(a) are all set to the periodic boundary conditions. In this way, we connect the tops of all the pillars to form a simple layered structure: the T-shaped steel on substrate form a cover layer. The geometric parameters are: $a$ = 1.5 m, $h$ = 6$a$, $d$ = 0.2$a$, $H$ = 500$a$ and $D$ = 0.25$a$. The bottom boundaries identified by the green lines are all set to the fixed boundary conditions.

Figure 11(b) shows the complex band structures of the surface modes of the TSM. Because the wavelength of Rayleigh waves at the ultra-low frequency (below 0.1 Hz) is too long to be difficult to calculate, only the complex band structures in the range of 0.1 to 8.0 Hz is drawn. *This is the first characteristic of the BG based on inverse-dispersion effect: there is no surface mode in the range of the BG.*

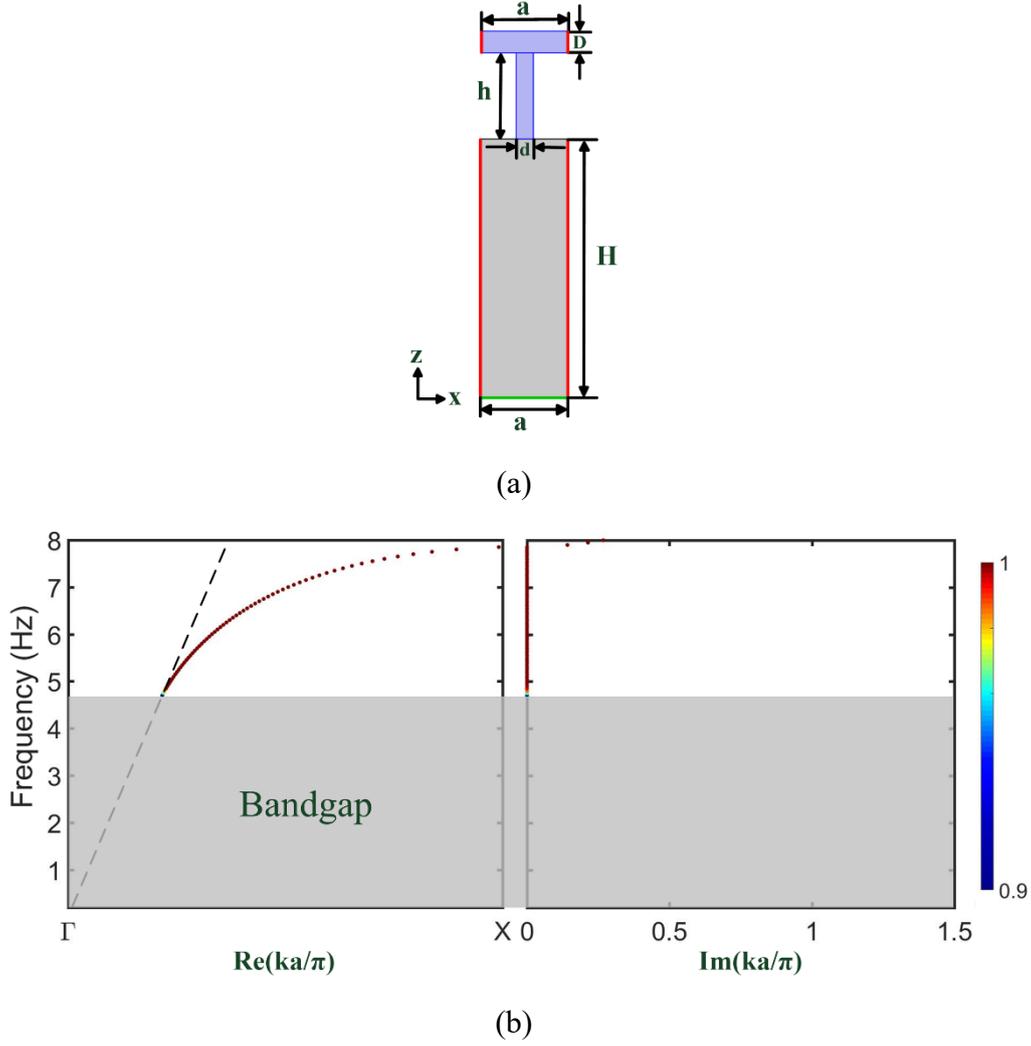

Figure 11: (a) The unit cell and (b) complex band structure of the TSM based on layer structure.

The TSM can be roughly equivalent to a layered medium composed of two materials as shown in Fig. 9. Considering that the cover layer is a steel structure composed of pillars and a plate, we calculate the effective Young's modulus and effective density of this cover layer at a low frequency (0.1 Hz) through effective medium theory [47,48]: Young's modulus $E$ = 207 GPa, density $\rho$=4000 kg/m³. Considering that the materials of the plates and pillar are all steel, the effective Poisson's ratio of this cover layer is still 0.3. And the thickness of the cover layer is $h_1$ = 6.25$a$, i.e. the height of pillar $h$ plus thickness of steel plate *D*. This layered medium has only a single guided wave mode and there is a cutoff frequency of about 0.0023 Hz in Fig. 12. In this layered medium, when the frequency is higher than 0.0023 Hz, there is no longer any Rayleigh wave. This is the reason for the BG shown in Fig. 11(b). This phenomenon has also been found in the SM with a similar structure.[26] *This is the second characteristic of the BG based on inverse-dispersion effect: SM can be equivalent to a two-layered medium, and it satisfies $V_{S1}>V_{r1}>V_{S2}>V_{r2}$.*

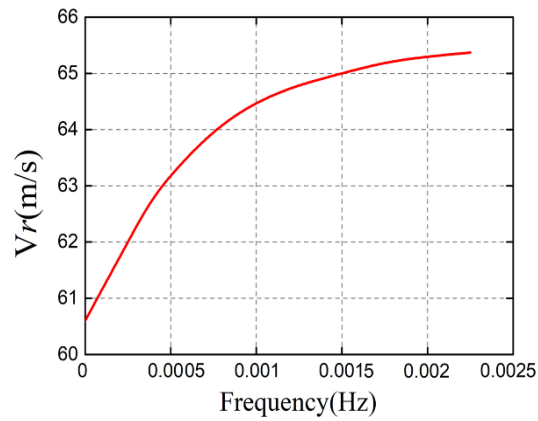

Figure 12: The dispersion curves of Rayleigh waves of the effective two-layered medium of the TSM.

The frequency domain analysis of the TSM is conducted to verify the existence of the BG, according to the model of the finite system shown in Fig. 13(a). Because the ultra-low frequency Rayleigh waves has a longer wavelength, the parameters we set are larger when designing the finite system to ensure the accuracy of the simulation results. Here, the number of unit cells of the TSM is set to 5000, mainly for lower frequency Rayleigh waves, such as below 0.1 Hz. The problem of the number of unit cells of TSM will be discussed in detail later. As shown in Fig. 13(b), we plot the displacement of the surface of this finite system at different frequencies. The TSM is placed at 0 < X < 7500

m. From Fig. 13(b), we can find that the TSM can attenuate Rayleigh waves well in the wide frequency range of 0.1 - 4.8 Hz.

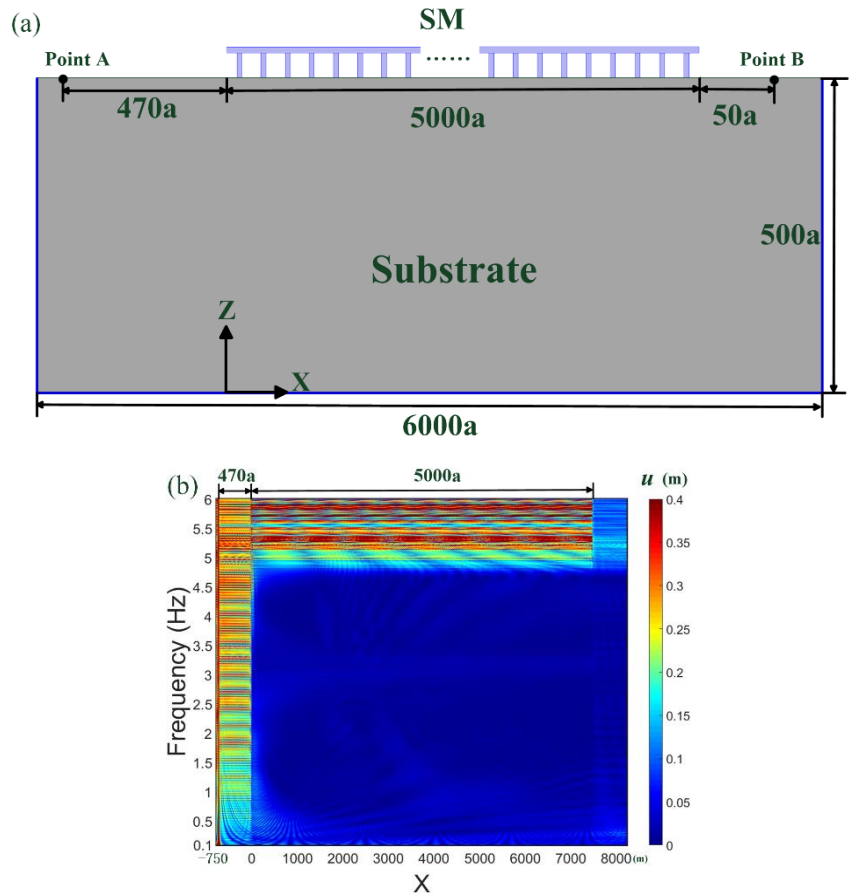

Figure 13: (a) A schematic diagram of the finite system for calculating transmission spectra. Note that the purple and gray areas represent steel and soil, respectively. (b) Distribution of displacement amplitude of the finite system varying with the frequency.

In order to discuss the effect of the number of unit cells of the TSM on the attenuated Rayleigh waves, the attenuation of the Rayleigh waves by the TSM with 200 unit cells is calculated. For a slightly lower frequency, such as 1.3 Hz, the TSM with 200 unit cells can convert a portion of the Rayleigh waves into a bulk waves, but the rest of the Rayleigh wave also passes through the TSM in Fig. 14(a). Obviously, the attenuation effect of the TSM with 200 unit cells is limited. As shown in Fig. 14(b), the TSM with 5000 unit cells are used can well convert most of the Rayleigh waves into bulk waves. At a slightly higher frequency, such as 3.5 Hz in Fig. 14(c-d), the attenuation effect of the Rayleigh waves by the TSM of 200 and 5000 unit cells is almost the same. Rayleigh waves are almost converted into bulk waves by the TSMs. *This is the third characteristic of the BG based on inverse-dispersion effect: the SM with enough unit cells can convert most of surface waves to bulk waves.*

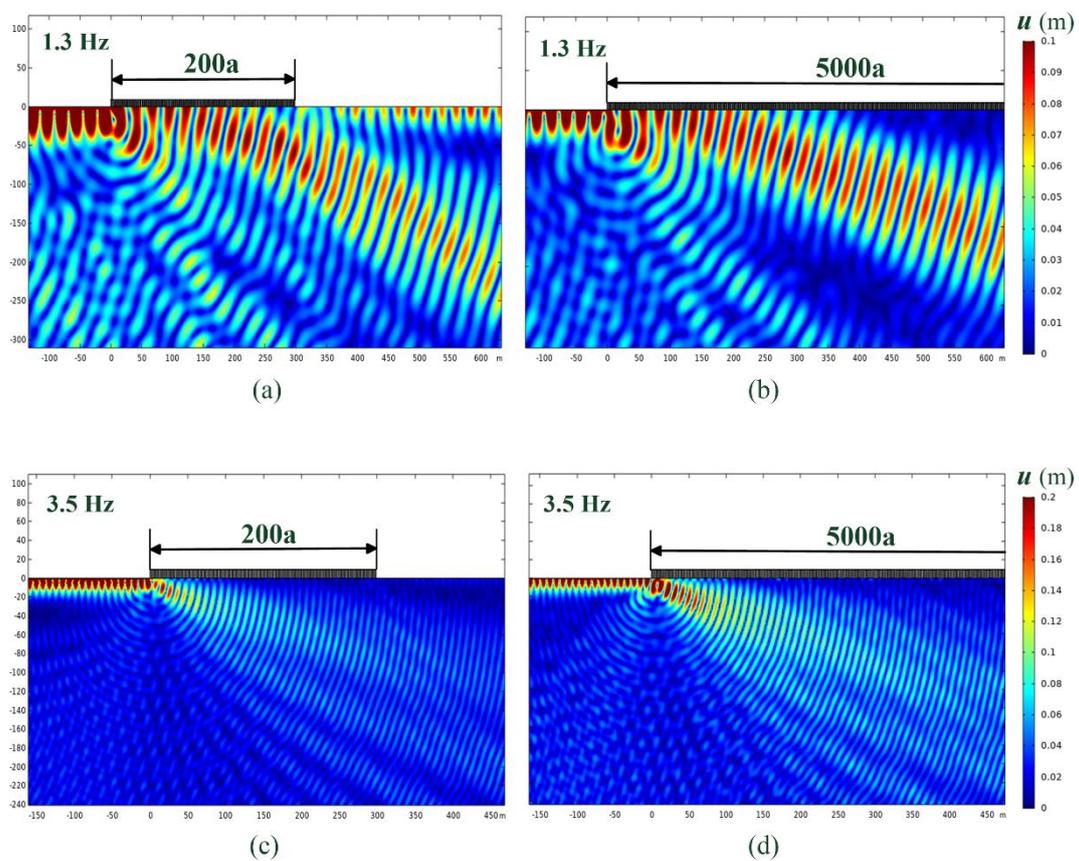

Figure 14: Distribution of displacement amplitude at 1.3 Hz for TSM with (a) 200 and (b) 5000 unit cells and at 3.5 Hz for (c) 200 and (d) 5000 unit cells.

**Appendix C: Propagation of Rayleigh waves without the 2D ITSM**

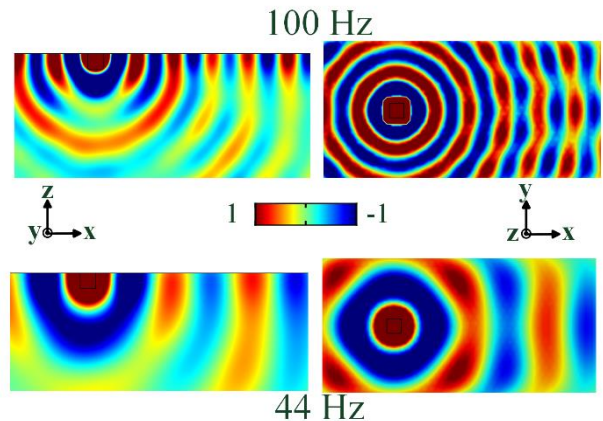

Figure 15: The simulated results of propagation of Rayleigh waves without the 2D ITSM at 44 and 100 Hz, respectively.